\documentclass[12pt]{article}
\usepackage{graphicx,amsmath,amssymb,mathrsfs}
\usepackage{epsfig}

\begin{document}

\begin{center}

{\bf Statistical mechanics of the vacuum}

\vspace{1cm}

{\bf Christian Beck}

\vspace{1cm}

{Queen Mary University of London, School of Mathematical Sciences, Mile End Road, London E1 4NS, UK}

\vspace{1cm}

\abstract{
The vacuum is full of virtual particles which exist
for short moments of time.
In this paper we
construct a chaotic model
 of vacuum fluctuations associated with a fundamental entropic field
 that generates an arrow of time.
 The dynamics can be physically interpreted in terms of fluctuating virtual momenta. This model leads to a generalized
statistical mechanics that distinguishes fundamental constants
of nature.}

\end{center}

\newpage

\section{Introduction}

Statistical mechanics has been applied to a large variety of
complex systems, with short-range and long-range
interactions and many different underlying microscopic dynamics.
Still what has never been done is to develop a kind of
statistical mechanics formalism for the vacuum
itself. Surely the vacuum, according to
Heisenberg's uncertainty principle, contains
a large number of virtual particles that exist for short
moments of time. Hence there is a large number
of fluctuating momentum and position variables,
which a priori should allow for a statistical mechanics description,
at least in a generalized sense.

In this paper we consider a simple model
that could be regarded as a first step
towards such a statistical mechanics description of
spontaneous momentum fluctuations in
the vacuum. On a microscopic level,
our approach will lead to a so-called
chaotic string dynamics
\cite{physicad,book,dettmann1,thomas,prd,dettmann2,groote1,groote2,
beck07,maher,groote3,schaefer1,schaefer2}. This is an additional dynamics
to the standard model field equations---basically the dynamics of an
entropy producing chaotic field.
This field is not directly measurable but evolves within the bounds
set by the uncertainty relation.

Our model of the vacuum is interesting for
two reasons: Firstly, an arrow of time is naturally produced
by this model, since the underlying chaotic dynamics corresponds
to a coupled Bernoulli shift of information which distinguishes
one direction of time. Secondly, certain observables
associated with chaotic strings have been previously
shown \cite{physicad,book} to distinguish standard model parameters as corresponding to
states of minimum vacuum energy. In that sense the
generalized statistical mechanics model considered
has the scope to fix fundamental constants by first principles.

\section{Fluctuations of momenta and positions}
Let us construct a probabilistic model of vacuum fluctuations.
Consider an arbitrary spatial direction in 3-dimensional empty
space, described by a unit vector $\vec{u}$.
%Assuming that
%superstring or M-theory is the correct theory to describe
%nature, then $\vec{u}$ may also point into a direction spanned up
%by the extra dimensions.
We know that the vacuum is full of
virtual particle-antiparticle pairs, for example $e^+e^-$ pairs.
Consider virtual momenta of such particles, which exist for short
time intervals due to the uncertainty relation. From quantum
mechanics (indeed, already from 1st quantization) we know that the
phase space is effectively divided into cells of size of
$O(\hbar)$. The uncertainty principle implies
\begin{equation}
\Delta p \Delta x =O(\hbar ).
\end{equation}
Here $\Delta p$ is some momentum uncertainty in $\vec{u}$
direction, and $\Delta x$ is a position uncertainty in $\vec{u}$
direction. In the following, we will regard $\Delta p$ and $\Delta
x$ as random variables. A priori we do not know anything about the
dynamics of these random variables. But three basic facts are
clear:

\begin{itemize}
\item[(i)]
$\Delta p$ and $\Delta x$ are not independent. Rather, they are strongly
correlated: A large $\Delta p$ implies a small $\Delta x$, and vice
versa.

\item[(ii)]
There are lots of virtual particles in the vacuum. For each of
those particles, and for each of the directions of space, there is
a momentum uncertainty $\Delta p^i$ and a position uncertainty
$\Delta x^i$, making up a phase space cell $i$.

\item[(iii)]
If the particle of a virtual particle-antiparticle pair has momentum
$\Delta p^i$, then the corresponding antiparticle has momentum
$-\Delta p^i$ in the rest frame.

\end{itemize}

We label the phase
space cells in $\vec{u}$-direction
by the discrete lattice coordinate $i$, taking values in $\bf{Z}$.
With each phase space cell $i$ we associate a rapidly fluctuating
scalar field
$\Phi_n^i$. It represents the momentum uncertainty
$\Delta p = p_n^i$ in $\vec{u}$ direction in cell $i$ at time
$t=n\tau$ in units of some maximum momentum $p_{max}$:
\begin{equation}
p_n^i = p_{max} \; \Phi_n^i \;\;\;\;\;\;\; \Phi_n^i \in [-1,1]
\end{equation}
$\tau$ is some appropriate time unit.
%Whether $t$ is physical time
%or fictitious time is irrelevant at the moment--- we just regard
%it as some suitable time variable.
We may call $p_n^i$ the
`spontaneous momentum' associated with cell $i$ at time $n$. The
particle and antiparticle associated with cell $i$ have momentum
$p_n^i$ and $-p_n^i$.

Let us introduce a position uncertainty variable $x_n^i$ by
writing
\begin{equation}
p_n^i x_n^i = \Gamma^{-1} \hbar \label{almut4}
\end{equation}
Here $\Gamma$ is a constant of order 1.
Indeed, eq.~(\ref{almut4}) states that the position
uncertainty random variable $x_n^i$ is essentially
the same as the inverse momentum uncertainty
random variable $1/p_n^i$, up to a constant times $\hbar$.
By this we certainly realize property (i).
The variable $x_n^i$ is related to the
rapidly fluctuating field $\Phi_n^i$ by
\begin{equation}
x_n^i = \frac{\hbar}{\Gamma p_{max}} \frac{1}{\Phi_n^i}
\end{equation}
By definition, the sign of
$x_n^i$ is equal to the sign of the field $\Phi_n^i$.

The phase space cells, labeled by the index $i$, can be
represented as intervals of constant length $\hbar / \Gamma$. Each
phase space cell is 2-dimensional (1 momentum, 1 position
coordinate). Although the volume of the cells is constant, the
shapes of the phase space cells fluctuate rapidly, since (in
suitable units) their side lengths are given by $\Phi_n^i$ and
$1/\Phi_n^i$. Indeed, $p_n^i$ and $x_n^i$ fluctuate both in time
$n$ and in the lattice direction $i$, and are uniquely determined
by the random variable $\Phi_n^i$.

It is clear that the $\Phi_n^i$ must have strong stochastic
or chaotic properties in order to serve as a good model for vacuum
fluctuations. On the other hand, since no physical measurements are able
to determine the precise momentum and position within a phase space cell
(due to the uncertainty relation), the dynamics of the $\Phi_n^i$ is
a priori unknown. We cannot measure a concrete time sequence of
vacuum fluctuations. Nevertheless, we are able to measure {\em
expectations} of vacuum fluctuations.

\section{Newton's law and self interaction}
Let us now introduce a dynamics for the field $\Phi_n^i$. It will
ultimately lead to a coupled map lattice,
the so-called chaotic string dynamics
\cite{physicad,book}, which is underlying
our statistical mechanics of the vacuum at a microscopic
level.
It should
be clear that the dynamics itself is not observable, due to the
uncertainty relation, but that expectations with respect to the
dynamics should be measurable in experiments. We attribute to each
phase space cell $i$ a self-interacting potential that generates
the dynamics. For example, we may choose a $\phi^4$-theory, where
the potential is of the form
\begin{equation}
V (\Phi) = \left( \frac{1}{2} \mu^2 \Phi^2 + \frac{1}{4} \lambda
\Phi^4 \right) mc^2 +C.
\end{equation}
In our physical interpretation $\Phi$ is dimensionless, $m$ is
of the order of the mass of the virtual particles under
consideration, $\mu^2$ and $\lambda$ are
dimensionless parameters, and $C$ is an additive constant. The
`force' in $\vec{u}$-direction due to this self-interacting
potential is given by
\begin{equation}
F (\Phi )= - \frac{1}{c\tau} \frac{\partial}{\partial \Phi} V (\Phi)
=(-\mu^2 \Phi -\lambda \Phi^3 ) \frac{mc}{\tau}
\end{equation}
(the factor $1/(c\tau )$ is needed for dimensional reasons,
regarding $t$ as physical time). We assume that the change of
momentum is given by Newton's law
\begin{equation}
\frac{\partial}{\partial t}p = p_{max} \; \dot{\Phi} = F (\Phi) .
\label{almut}
\end{equation}
Note that Newton's law is also valid in a relativistic setting, provided $t$
denotes proper time.

The smallest time unit of our model of vacuum fluctuations is
$\tau$. This means that it does not make sense to consider
Newton's law for infinitesimally small time differences $\Delta
t$, since these would yield infinite energies $\Delta E$, from
$\Delta E \Delta t = O(\hbar )$. Thus we write eq.~(\ref{almut})
in the finite-difference form
\begin{equation}
p_{max} \frac{\Phi_{n+1} -\Phi_n}{\tau} =( -\mu^2 \Phi_n -\lambda \Phi_n^3)
\frac{mc}{\tau}.
\end{equation}
It is remarkable
that the unknown time lattice constant $\tau$ drops
out. For arbitrary $\tau$ we get a dynamics that is given by the
cubic mapping
\begin{equation}
\Phi_{n+1} = \left( 1-\frac{\mu}{\nu} \right) \Phi_n - \frac{\lambda}{\nu}
\Phi_n^3, \label{almut5}
\end{equation}
where $\nu := \frac{p_{max}}{mc}$. The dynamics of the
$\Phi$-field is time-scale invariant, it does not depend on the
arbitrarily chosen time lattice constant $\tau$.

We can obtain a cubic mapping of type (\ref{almut5})
in the following two very different
situations. Either $\nu =O(1)$, i.e. $p_{max} =O(mc)$, and the
potential parameters $\mu^2$ and $\lambda$ are of $O(1)$ as well.
In this case we consider a low-energy model of vacuum fluctuations.
We can, however, also let $p_{max} \to \infty$, thus considering
a high-energy theory. In this case $\nu =\frac{p_{max}}{mc} \to \infty$.
However, we can get {\em the same} finite cubic mapping if at the same time
the parameters $\mu^2$ and $\lambda$ of the self-interacting
potential diverge
such that $\mu/\nu$ and $\lambda /\nu$ remain finite.
%This is just the anti-integrable limit mentioned before.

There are distinguished parameter values leading to
Tchebyscheff maps and thus to strongest possible chaotic
behaviour. The negative 3rd-order Tcheby\-scheff map
$\Phi_{n+1}=3\Phi_n-4\Phi_n^3$ is obtained
from the potential
\begin{equation}
V_-^{(3)} (\Phi ) = \nu (-\Phi^2 +\Phi^4 )mc^2 +C_- , \label{pot1}
\end{equation}
the corresponding force is
\begin{equation}
F_-^{(3)}(\Phi ) = \nu (2\Phi -4 \Phi^3 ) \frac{mc}{\tau}.
\end{equation}
The positive 3rd-order Tchebyscheff map $\Phi_{n+1}=
-3\Phi_n+4\Phi_n^3$ is obtained from
\begin{eqnarray}
V_+^{(3)} (\Phi ) &=& \nu (2\Phi^2 -\Phi^4 )mc^2 +C_+
\label{pot2}
\\ F_+^{(3)} (\Phi ) &=& \nu (-4\Phi +4\Phi^3 ) \frac{mc}{\tau}.
\end{eqnarray}
In fact, we can get any Tchebyscheff map $\pm T_N$ of order $N$,
by considering appropriate potentials \cite{book,prd}.

Note that the strength of the potential is dependent on the energy
scale $E_{max}=p_{max}c$ at which we look at the vacuum. This is
indeed reasonable for a model of vacuum fluctuations. Namely, the
potential should be proportional to the energy $\Delta E$
associated with a vacuum fluctuation, and from $\Delta E \Delta t
=O(\hbar)$ we expect a larger energy on a smaller scale. Still the
form of the dynamics in units of $E_{max}$ is scale invariant. If
a small coupling between neighbored phase space cells is
introduced, approximate scale invariance is still retained. This
is similar to velocity fluctuations in a fully developed
turbulent flow, which are also approximately scale invariant
and strongly chaotic.

We note that a Tchebycheff dynamics $T_N$
with $N\geq 2$ produces information
in each iteration step \cite{schloegl}. It is
conjugated to a Bernoulli shift of $N$ symbols. There is no time-reversal
symmetry since each iterate
$\Phi_n$ has $N$ pre-images. Hence
our chaotic dynamics describing vacuum fluctuations is something new,
something in addition to the standard model field equations
which do not have an arrow of time.
We may
associate the dynamics with a fundamental entropic field, whose
main role is to produce information and to distinguish a particular
direction of time.

%Note that any Tchebyscheff dynamics
%\begin{equation}
%\Phi_{n+1}^i = \pm T_N(\Phi_n^i)
%\end{equation}
%is distinguished in the sense that it can be regarded as a
%`rotation' of an angle $u$ with `exponential acceleration'
%\begin{equation}
%\Phi_n^i = \pm \cos \pi  N^n u^i \;\;\;\;\;\;\;\; (u^i \in [0,1]).
%\end{equation}
%This somewhat reminds us of a classical de-Sitter state with
%exponential expansion, viewed stroboscopically.

\section{Coulomb forces and Laplacian coupling}

We may associate the strongly fluctuating variables $p_n^i$ with
the momenta of charged virtual particles
that are created out of the self energy of the entropic field.
 For example, we may
think of electrons and positrons, or any other types of fermions.
Actually, we should think of a collective system of such charged
particles, similar to a Dirac lake.

%The following consideration will make it plausible why there
%should be 1-dimensional Laplacian ($=$ diffusive) coupling between
%the fluctuating momenta. We will see that within our many-particle
%interpretation the coupling constant $a$ is related to the
%strength of an $1/r$-potential.

Suppose that (for example) $\Phi_n^i$ represents the momentum of
a virtual electron and $\Phi_n^{i+1}$ the momentum of a neighbored
virtual positron. The Coulomb potential between two opposite charges at
distance $r=|\vec{r}|$ is
\begin{equation}
V_{el} (r) =-\hbar c \alpha \frac{1}{r} .
\end{equation}
$\alpha\approx 1/137$ is the fine structure constant. The force (=
momentum exchange per time unit $\Delta t$) is
\begin{eqnarray}
F_{el}(r) = \frac{\Delta p_{el}}{\Delta t}
& =& - \frac{\partial}{\partial r} V_{el} (r)
 = -\hbar c\alpha \frac{1}{r^2}
  \\
\Longleftrightarrow \Delta p_{el} &=& -\hbar c\alpha \frac{\Delta
t}{r^2}.
\end{eqnarray}
Again, due to the uncertainty relation $\Delta E \Delta t =O(\hbar )$
it does not make sense to choose an
infinitesimally small time unit $\Delta t$. It is more reasonable to choose
\begin{equation}
c \Delta t =r
\end{equation}
since photons move with the velocity of light.
We then end up with the fact that the Coulomb potential gives rise to
the momentum transfer
\begin{equation}
\Delta p_{el} = -\hbar \alpha \frac{1}{r} \label{almut15}
\end{equation}
during the time unit $\Delta t =r/c$.

In our picture of vacuum fluctuations, distances $\Delta x$ and
hence also inverse distances $1/r$ are strongly fluctuating due to
the uncertainty relation. In section 2 we attributed the
strongly fluctuating inverse distance variable
$|\frac{\Gamma}{\hbar} p_n^i|$ to each phase space cell $i$. The
maximum value of the inverse distance, corresponding to the
smallest possible distance at a given energy scale $p_{max}$, is
given by $\frac{\Gamma}{\hbar} p_{max}$. What should we now take
for the inverse {\em interaction} distance between two neighbored
particles $i$ and $i+1$? Obviously, the relevant quantity is the
momentum {\em difference} between them. According to the
uncertainty relation, local momentum differences always correspond
to local inverse distances between cells. Hence we define the
inverse interaction distance $\frac{1}{r}_{i,i+1}$ between the
electron in cell $i$ and the positron in cell $i+1$ as the
following strongly fluctuating random variable:
\begin{equation}
\frac{1}{r}_{i,i+1}
=\frac{\Gamma p_{max}}{2\hbar} |\Phi_n^{i+1} -\Phi_n^i|  \label{almut6}
\end{equation}
The factor $\frac{1}{2}$ is needed to let the inverse interaction
distance not exceed the largest possible value
$\frac{\Gamma}{\hbar} p_{max}$ of the inverse distance. It follows
that the absolute value of the momentum transfer between cell $i$
and cell $i+1$ is
\begin{equation}
|\Delta p_{i,i+1} | = \hbar \alpha \frac{1}{r}_{i,i+1} = p_{max}
\Gamma \frac{\alpha}{2} |\Phi_n^{i+1} -\Phi_n^i |
\end{equation}
The momentum transfer can be positive or negative with equal
probability, depending on whether we have equal or opposite
charges in the neighbored cells. A possible choice for the signs
is to take
\begin{equation}
\Delta p_{i,i+1}  = p_{max}\Gamma \frac{\alpha}{2} (\Phi_n^{i+1}
-\Phi_n^{i}) \label{almut66}
\end{equation}
This, indeed, causes
inhomogeneities of the $\Phi$-field to be smoothed out:
If $\Phi^{i+1} > \Phi^i$,
$\Phi^i$ increases. If $\Phi^i =\Phi^{i+1}$
(homogeneity), there is no change at all. If $\Phi^{i+1} < \Phi^i$,
$\Phi^i$ decreases.

Similarly, the momentum transfer from the left neighbor is
\begin{equation}
\Delta p_{i,i-1} =p_{max} \Gamma \frac{\alpha}{2} (\Phi_n^{i-1}
-\Phi_n^i )
\end{equation}
Thus
\begin{equation}
\Delta p_{i+1,i} =p_{max} \Gamma \frac{\alpha}{2} (\Phi_n^i
-\Phi_n^{i+1} ) = - \Delta p_{i, i+1}.
\end{equation}
The momentum exchange is antisymmetric under the exchange of the two particles,
as it should be. We now have a direct physical interpretation
associated with the signs of the $\Phi$-field: If $\Phi_n^{i+1} > \Phi_n^i$,
we have opposite charges in cell $i$ and $i+1$, causing attraction.
Otherwise, the charges have equal signs, causing repulsion.

The above approach corresponds to diffusive coupling.
However, we
could also look at a momentum transfer given by $-p_{max}\Gamma
\frac{\alpha}{2} (\Phi_n^{i-1}+\Phi_n^i)$, as generated by
anti-diffusive coupling. In this case one assumes that the average
momentum $\frac{1}{2}(\Phi_n^{i-1}+\Phi_n^{i})p_{max}$ determines
the inverse interaction distance $1/r_{i-1,i}$. Note that this
approach still generates diffusive coupling if at the same time
one electron is re-interpreted as a positron, which formally,
according to Feynman, has a momentum of opposite sign. So both
coupling forms can be physically relevant.

In total, the momentum balance equation for cell $i$ is
\begin{eqnarray}
p_{n+1}^i &=& p_n^i +\Delta p_{i,i-1} + \Delta p_{i,i+1} \\
\Longleftrightarrow \Phi_{n+1}^i &=& \Phi_n^i +\Gamma
\frac{\alpha}{2} (\pm \Phi_n^{i-1} -2\Phi_n^i \pm \Phi_n^{i+1}),
\label{almut7}
\end{eqnarray}
where the $\pm$ sign corresponds to diffusive or anti-diffusive
coupling, respectively. Remarkably, the momentum cutoff $p_{max}$
drops out, and we end up with an evolution equation where only
dimensionless quantities $\Phi ,\alpha$ and $\Gamma$ enter. Notice
that eq.~(\ref{almut7}) is a discretized diffusion equation with
diffusion constant $\Gamma \alpha$. Also notice that $\Gamma$ is
just the constant of $ O(1)$ in the uncertainty relation.
$\Gamma^{-1}$ is the size of the phase space cells in units of
$\hbar$. If we {\em define} phase space cells to have size
$\hbar$, this implies $\Gamma =1$. Then the only relevant constant
remaining is the coupling strength $\alpha$ of the
$1/r$-potential.

Finally, we have to combine the chaotic self-interaction with
the diffusive interaction. Since, due to the uncertainty
principle, our time variable is effectively discrete, both
interactions must alternate. First, in each cell $i$ there is a
`spontaneous' creation of momentum due to the self-interacting
potential $V_\pm^{(N)}$:
\begin{equation}
\Phi_{n+1}^i = \pm T_N(\Phi_n^i)  . \label{almut9}
\end{equation}
Then, momenta of neighboured
particles smooth out due to Coulomb interaction.
Setting $\Gamma = 1$ we have
\begin{equation}
\Phi_{n+2}^i = \Phi_{n+1}^i +\frac{\alpha}{2} (\pm
\Phi_{n+1}^{i-1} -2 \Phi_{n+1}^i \pm \Phi_{n+1}^{i+1}).
\label{almut10}
\end{equation}
Combining eqs.~(\ref{almut9}) and (\ref{almut10}) we obtain the coupled
map lattice
\begin{equation}
\Phi_{n+2}^i =(1-\alpha) T_N(\Phi_n^i) \pm
\frac{\alpha}{2}(T_N(\Phi_n^{i-1}) +T_N(\Phi_n^{i+1}), \label{chastri}
\end{equation}
where $T_N$ can be either the positive or negative Tchebyscheff
map.
What we obtained by our simple
intuitive arguments is just the chaotic string dynamics
introduced in \cite{physicad,book,prd}
but now derived in a pedestrian, easy-going way.
Our physical derivation implies that
the coupling constant $\alpha$ can be identified with a
standard model coupling constant.

%Certainly we can also obtain backward coupling rather than forward
%coupling. This just means that the Coulomb interaction does not
%take place simultaneously at all lattice sites, but alternates
%with the self interaction in $i$ direction. All these chaotic
%string theories with their various degrees of freedom represent
%possible models of vacuum fluctuations. Our current formulation
%emphasizes phase space rather than space and deals with a large
%number of virtual particles. It has thus similarities with a
%thermodynamic description of the vacuum (for more details of this
%the next sections).
%In contrast to conventional statistical
%mechanics, where the momentum and position of a particle can be
%chosen independently, in our model the momentum uncertainty and position
%uncertainty are not independent from each other but connected
%through the uncertainty relation. Notice that our interpretation
%requires 1-dimensional lattices, since the uncertainty relation is
%valid for each space direction in an independent way.
%The Coulomb potential just changes the momentum component in one
%direction, the radial $\vec{u}$-direction between the particles.

\section{Feynman webs}

Let us now further work out our interpretation and proceed to a
more detailed physical interpretation of the chaotic string
dynamics. Remember that in this interpretation we regard
$\Phi_n^i$ to be a fluctuating momentum component associated with
a particle $i$ at time $n$.
Neighbored particles $i$ and $i-1$ exchange
momenta due to the diffusive coupling.

A more detailed physical interpretation would be that at each time
step $n$ a fermion-antifermion pair $f_1,\bar{f}_2$ is being
created in cell $i$ by the field energy of the self-interacting
fundamental entropic field.
 In units of some arbitrary energy scale $p_{max}$, the
fermion has momentum $\Phi_n^i$, the antifermion momentum
$-\Phi_n^i$. They interact with particles in neighbored cells by
exchange of a gauge boson $B_2$, then they
annihilate into boson $B_1$ and the next chaotic vacuum
fluctuation (the next creation of a particle-antiparticle pair)
takes place. This can be symbolically described by the Feynman
graph in Fig.~1.
\begin{figure}
\includegraphics[scale=1.0]{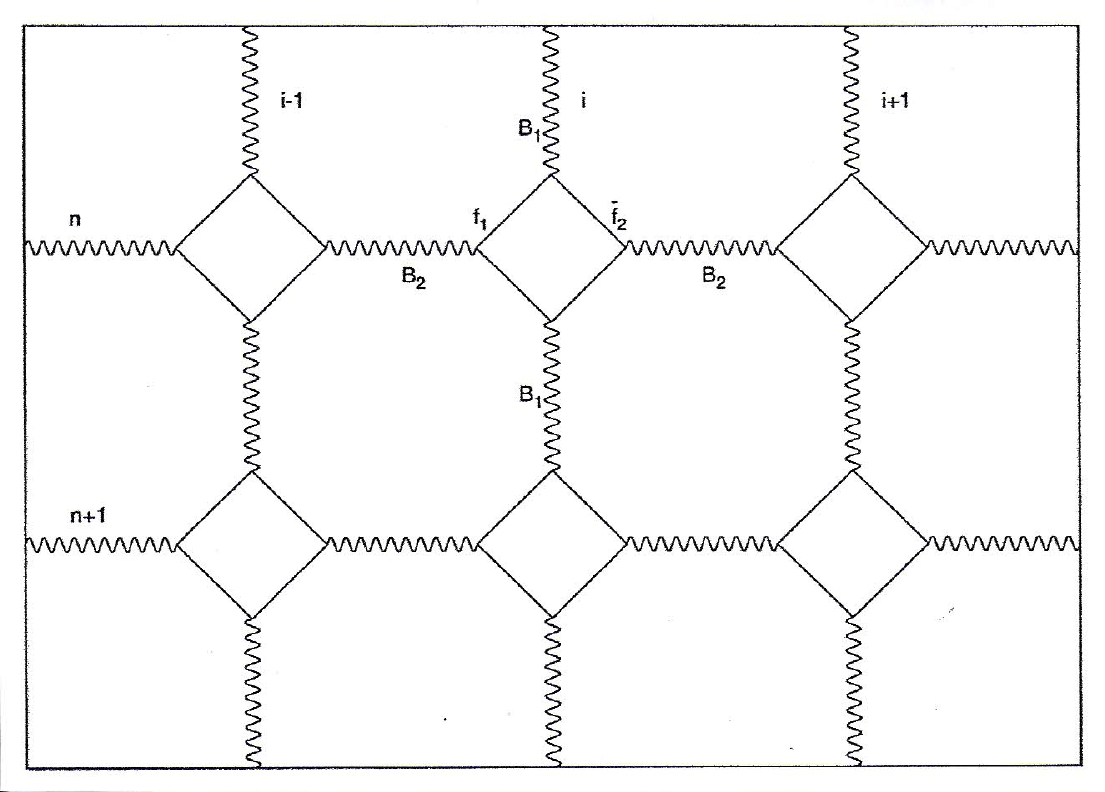}
 \caption{Feynman web
interpretation of the coupled map dynamics.}
\end{figure}
Actually, the graph continues ad infinitum in time and space and
could thus be called a `Feynman web', since it describes an
extended spatio-temporal interaction state of the
entropic field, to which
we have given a standard model-like interpretation. The important
point is that in this interpretation $\alpha$ is a
standard model coupling constant, since it describes the strength
of momentum exchange of neighbored particles.
%At the same time,
%$a$ can also be regarded as an inverse metric in the 1-dimensional
%string space, since it determines the strength of the Laplacian
%coupling.

It is well known that standard model interaction strengths
actually depend on the relevant energy scale $E$. We have the
running electroweak and strong coupling constants \cite{kane}.
For example,
the fine structure constant $\alpha_{el}(E)$ slightly increases
with $E$, and the strong coupling $\alpha_s$ rapidly decreases
with $E$.
What should we now take for the energy (or temperature) $E$ of the
chaotic string? In
\cite{physicad,book} extensive numerical evidence
was presented that
minima of the vacuum energy of the chaotic strings are observed
for certain distinguished string couplings $\alpha_i$, and these string
couplings are numerically observed to coincide with running
standard model couplings, the energy (or temperature) being given
by
\begin{equation}
E= \frac{1}{2} N (m_{B_1}+m_{f_1}+m_{f_2}). \label{katharina}
\end{equation}
Here $N$ is the index of the Tchebyscheff map of the chaotic
string theory considered, and $m_{B_1}, m_{f_1},m_{f_2}$ denote
the masses of the particles involved in the Feynman web
interpretation. The surprising observation is that rather than
yielding just some unknown exotic physics, the chaotic string
spectrum appears to reproduce the masses and coupling constants of
the known quarks, leptons and gauge bosons of the standard model
(plus possibly more).

%\footnote{If the presence of black holes is assumed, one can also
%relate $E$ to the Hawking temperature $kT_H\sim  \frac{1}{GM}$,
%thus connecting very large black hole masses $M$ with very small
%temperatures $kT_H$.}.
Formula (\ref{katharina}) formally  reminds us of the
zeropoint energy
levels $E_N=\frac{N}{2} \hbar \omega$ of $N$ quantum mechanical
harmonic oscillators.
In the Feynman web
interpretation of Fig.~1, the formula is plausible. We expect
the process of Fig.~1 to be possible as soon as the energy per
cell $i$ is of the order $m_{B_1}+m_{f_1}+m_{f_2}$. The boson
$B_2$ is virtual and does not contribute to the energy scale. The
factor $N$ can be understood as a multiplicity factor counting the
number of degrees of freedom. Given some value $\Phi_n^i$ of the
momentum in cell $i$, there are $N$ different pre-images
$T_N^{-1}(\Phi_n^i )$ how this value of the momentum can be
achieved. All these different channels contribute to the energy
scale.

\section{Heat bath of the vacuum}
Let us now work out a statistical mechanics of the vacuum in
somewhat more detail.
We regard the vacuum as a kind of heat bath of virtual
particles. Generally we want to use
concepts from statistical mechanics and information theory. First,
consider ordinary statistical mechanics. Given a system of $N_p$
classical particles with Hamiltonian
\begin{equation}
H=\sum_{i=1}^{N_p} \frac{\vec{p}_i^{\; 2}}{2m_i} +\frac{1}{2}
\sum_{i,j} V(\vec{q}_i,\vec{q}_j)
\end{equation}
the probability density $\rho$ to observe a certain microstate
$(\vec{q}_1,\ldots ,\vec{q}_{N_p},$ $ \vec{p}_1, \ldots$
$\ldots ,\vec{p}_{N_p})$ $:=(q,p)$ is
\begin{equation}
\rho (q,p)=\frac{1}{Z(\beta)}e^{-\beta H(q,p)}.
\end{equation}
We assume that Boltzmann statistics is applicable.
$\beta=1/kT$ is the inverse temperature and
\begin{equation}
Z(\beta ) =\int d\vec{q_1} \cdots d\vec{q}_{N_p}d\vec{p_1} \cdots
d\vec{p}_{N_p} e^{-\beta H(q,p)}
\end{equation}
is the partition function. The internal energy $U$ is defined as
the expectation of $H$:
\begin{equation}
U=\langle H \rangle =\int d\vec{q}_1 \cdots
d\vec{q}_{N_p}d\vec{p}_1 \cdots d\vec{p}_{N_p}
 \rho (q,p) H (q,p) = - \frac{\partial}{\partial \beta} \log Z(\beta )
\label{interna}
\end{equation}

If we want to develop a thermodynamic description of the vacuum we
need a statistical theory of vacuum fluctuations $\Delta q_i$ and
$\Delta p_i$ allowed by the uncertainty relation. The situation,
however, is different from ordinary statistical mechanics because
the momentum and position variables $\Delta q_i, \Delta p_i$
cannot be chosen independent from each other, as for ordinary
statistical mechanics. If we choose a certain $\Delta p_i$ then
$\Delta q_i=\hbar /\Delta p_i$ is already fixed. Moreover, since
virtual momenta violate energy conservation (they are just defined
as doing that), we cannot expect to have an ordinary Hamiltonian
$H(q,p)$ as in classical mechanics. If anything, the dynamics
should be dissipative. This is why we base our statistical
mechanics of the vacuum on the dissipative dynamics
of section 2-4.

We also have to decide what `temperature' means for the vacuum. It
appears most reasonable to identify $kT\sim E$ with the energy
scale $E$ at which we look at the vacuum. Then
$q_{min}=O(\frac{\hbar c}{kT})$ is the smallest spatial scale
resolution we can achieve at this temperature, and
$p_{max}=\frac{\hbar}{q_{min}}$ is the maximum momentum. As worked
out in the previous sections, the relevant information on the state of a phase
space cell of size $\hbar$ is assumed to be given by a field
variable $\Phi_n^i$, which is the momentum uncertainty $\Delta
p_i=p_n^i$ in units of $p_{max}$ at time $n$ in cell $i$. The
corresponding position uncertainty is $x_n^i=\hbar/ p_n^i$.

Since the vacuum is isotropic, the direction in which we measure
the momentum is irrelevant. If there are $d$ spatial directions,
the $d$ components $\Delta p_{x_1}, \ldots , \Delta p_{x_d}$ of
the momentum uncertainty into the $d$ space directions are
expected to be independent from each other. In empty space, we do
not expect any interactions between $\Delta p_{x_1}$ and $\Delta
p_{x_2}$ for two different directions. Rather, 1-dimensional
models are expected to do a good job.

We can either construct models where $\Phi_n^i$ is a pure random
field, or where there is an underlying chaotic dynamics. The second
type of models, which leads to chaotic strings in a natural way,
has been shown \cite{physicad,book,prd,beck07,maher} to reproduce
observed standard model parameters
(fermion and boson masses, coupling constants,
mixing angles) with very
high precision, taking as the leading principle the minimization
of vacuum energy.

We do not have a
true Hamiltonian for the chaotic string dynamics since the dynamics is
dissipative. But we can write down a kind of
analogue of a Hamiltonian given by
\begin{equation}
H = \sum_i V_\pm (\Phi^i) +aW_\pm (\Phi^i,\Phi^{i+1}),
\end{equation}
with the self-interacting potential $V$
given by eq.~(\ref{pot2}) (or its generalization)
acting first, then followed by
a potential $W\sim (\Phi^i-\Phi^{i+1})^2$ generating the diffusive coupling via nearest
neighbor interaction.
Due to the uncertainty
relation, the time variable is effectively discrete with a lattice
constant of order $\Delta t=\hbar/ E$. In order to define an
internal energy of vacuum fluctuations similar to
eq.~(\ref{interna}) we thus have to decide whether we relate it to
$V$ at one time step, to $W$ at the next time step, or to the sum
of both averaged over both time steps. These degrees of freedom
are absent in classical statistical mechanics, where the time
evolution is continuous. All three types of vacuum energies
are important \cite{book}.

The equilibrium distributions, replacing the canonical probability
distributions of ordinary statistical mechanics, are the invariant
densities $\rho(\Phi^1 ,$ $ \Phi^2 , \ldots )$ of the coupled map
dynamics. In contrast to ordinary statistical mechanics, there is no
simple analytic expression for them, except for the uncoupled case
$\alpha=0$, where we have \cite{schloegl}
\begin{equation}
\rho (\Phi^1 , \Phi^2 , \ldots )= \prod_i \frac{1}{\pi \sqrt{1-{\Phi^i}^2}}.
\end{equation}
These types of densities can be dealt with in the formalism of
nonextensive statistical mechanics \cite{tsallis1, tsallis2},
they correspond to $q$-Gaussians with $q=3$, respectively $q=-1$
if the escort formalism is used \cite{schloegl,beckq}.
 Generally,
the invariant densities depend on the coupling $\alpha$ in a
non-trivial way. All averages are formed with these densities. For
ergodic systems the ensemble averages can be replaced by time
averages.

Note that a dynamics generated by a Tchebyscheff map does not have
a unique inverse, hence an arrow of time arises in a natural way.
This arrow of time of the heat bath of the vacuum helps to
justify the arrow of time in ordinary statistical mechanics,
it is associated with our fundamental entropic field.
Whereas classical mechanics is invariant under time reversal, the
dynamics of the vacuum fluctuations considered here is not.
In our approach the arrow of time enters at a fundamental
level, as a hidden entropic dynamics of the vacuum.

\section{States of maximum information and minimum correlation}

Given the potential $V$ of eq.~(\ref{pot2}), or its generalizations
discussed in \cite{prd}, the expectation
$\langle V(\Phi ) \rangle (a) =:V(a)$ measures the
self energy of the vacuum per phase space cell (or per virtual
particle). This can be regarded as a kind of thermodynamic potential
of the vacuum. Numerically it is obtained by iterating the coupled
map lattice (\ref{chastri}) for a given coupling
$\alpha =a$, choosing random initial conditions and
averaging $V_+^{(N)}(\Phi_n^i)$ over all $n$ and $i$ (disregarding
the first few transients).
% Another thermodynamic potential
%of interest is the interaction energy of nearest neighbours
%$\langle W(\Phi^i,\Phi^{i+1})\rangle =:W(g)$. Since the action of
%$V(\Phi^i)$ and $W(\Phi^i,\Phi^{i+1})$ alternates in time, both
%termodynamic potentials should be considered on an equal footing,
%they are of similar importance. Of course we can also look at the
%total vacuum energy $H(g):=\langle V(\Phi^i) +
%W(\Phi^i,\Phi^{i+1})\rangle$.
Numerical results for $V(a)$ were presented in detail in \cite{physicad, book}.
It turns out that this function
typically varies smoothly with $a$ but has lots of local minima
and maxima. What is the physical interpretation of such an
extremum?

We may interpret $V(a)$ as a
kind of entropy function of the vacuum. Clearly the Tchebyscheff
maps, as any chaotic maps, produce information when being
iterated. Or, looking at this the other way round, information on
the precise initial value is lost in each iteration step due to
the sensitivity on initial conditions
\cite{schloegl}. The potential $V(\Phi )$
generates the chaotic dynamics and hence could be formally
regarded as a kind of information potential. Its expectation
measures the missing information (=entropy) we have on the particle
contents of the phase space cells. At a minimum of $V(a)$ we have
minimum missing information. In other words, we have maximum
information on the particle contents of the cells. Hence we can
associate the dynamics with a particular Feynman web at this
point, and $a$ should then coincide with the corresponding
standard model coupling. This is what is indeed observed,
see \cite{physicad,book} for details.

Another interesting observable is the correlation function
$C(a)= \langle \Phi^i
\Phi^{i+1}\rangle$ of nearest neighbors.
One observes that
$C(a)$ typically varies smoothly with $a$, and that it vanishes at
certain distinguished couplings $a_i\not=0$
\cite{physicad,book,schaefer1}. States of the vacuum
with vanishing correlation
are clearly distinguished
--- they describe, in a sense, a state where the system, although
deterministic chaotic, is as random as possible.
A zero
of $C(a)$ means that the correlation between
the momenta of neighbored virtual particles vanishes, meaning that we can
clearly distinguish the particles in the various phase space
cells, so that again a Feynman web with a definite particle
contents makes sense. If a standard model coupling is chosen to
coincide with a zero of the interaction energy, then this clearly
represents a distinguished state of the heat bath of the vacuum,
with a vanishing spatial 2-point function just as for uncoupled
independent random variables, well suitable for stochastic
quantization methods \cite{parisi,huffel}.

%We also have to consider the stability of a certain state of the
%vacuum with zero interaction energy. Imagine that the coupling
%constant $a$ of the chaotic string changes slightly, then this
%change induces a change of the interaction energy of the vacuum.
%This change of the interaction energy is expected to influence the
%temperature $E\sim kT$, and this will once again induce another
%change of the running standard model coupling $a(E)$. If the
%coupling constant is driven back to its original value, then the
%state of the vacuum is stable.

%In the next chapter we will find straightforward physical
%interpretations for zeros where the correlation function has
%negative slope. Hence we conclude that only zeros with negative
%slope describe stable states of the vacuum where the coupling is
%driven to a stable equilibrium state.
%It appears that
%if $g$ is associated with a
%standard model interaction that is asymptotically free
%(such as strong interaction), the interaction energy
%$W(g)=\pm \frac{g}{2}C(g)$ must influence the temperature $E\sim kT$ with
%a negative sign, whereas
%for asymptotically non-free interactions (such as electric interaction),
%it must influence it with a positive sign, in order to achieve
%stability (see Fig.~10.1).

We could interpret the correlation function as describing the
polarization of the vacuum. Suppose, for example, that $\Phi_n^i$
represents a momentum component of an electron. Then, according to
Feynman, $-\Phi_n^i$ could be interpreted as the momentum
component of a positron. A negative correlation function $\langle
\Phi_n^i \Phi_n^{i+1} \rangle$ means that if there is an electron
in cell $i$, then with slightly larger probability there is a
positron in cell $i+1$, since the expectation of the product
$\Phi_n^i \Phi_n^{i+1}$ is negative. A zero of the correlation
function thus means the onset of vacuum polarization. Again we
expect the threshold points where vacuum polarization sets in to
occur at Feynman webs with energy $E=\frac{N}{2}kT$, with
$kT=m_{B_1}+m_{f_1}+m_{f_2}$. Numerical evidence that
stable zeros indeed coincide with running standard model
coupling constants evaluated at these energy scales has
been presented in \cite{physicad,book,schaefer1}.

\section{Conclusion and Outlook}

In this paper we have re-derived the chaotic string dynamics
previously introduced in \cite{physicad, book} in a way
that uses the language and tools of statistical mechanics.
We associated the rapidly fluctuating chaotic dynamics
with a fundamental entropic field that produces information
and that can
decay into virtual standard model particles. The description
then basically reduces to a novel statistical mechanics
description of the vacuum. The entropic field is responsible for the
arrow of time in nature at a fundamental level. Suitable
thermodynamic potentials can then be defined and investigated as a function
of the coupling parameter $a$. Numerical evidence
presented in \cite{physicad, book,
prd,groote1,beck07,groote3,schaefer1,schaefer2} has shown that these generalized
thermodynamic potentials distinguish standard model parameters.
Note that the dynamics underlying our approach
is discrete, nonlinear, chaotic, coupled
and complex. This is complexity science at a fundamental level.

An interesting question for future research is whether it is possible
to embed our statistical mechanics of the vacuum into some greater
theory.
Maybe the fundamental entropic field generating the
chaotic dynamics is something that is really relevant
before the creation of space-time, at a stage of the universe
where information (rather than matter and radiation) is the
relevant concept and standard model properties (including
fundamental constants) are still in the process of being fixed.
The space-time foam of string theory might yield another
embedding scenario \cite{xx}. At the Planck scale,
quantum coherence is likely to be broken due to spacetime foams
and the Hawking effect, and there is clearly
some sort of effective coarse graining going on at
this scale, so that tools from statistical mechanics
become more and more important.

Our derivation was based on
classical mechanics, but relativity can emerge out of classical mechanics
using superstatistical path integrals, as shown in \cite{yy}.
Intrinsically,
our statistical mechanics of the vacuum
can indeed be related to the superstatistics concept
\cite{beck-cohen}, in the sense that the vacuum
fluctuations of our model have non-Gaussian probability distributions
which can be obtained by averaging over many inverse temperatures.
These different inverse temperatures correspond to the various scales on which vacuum fluctuations take place. The superstatistical description
then involves a fundamental symmetry, the Euclidean group in
1 dimension, as shown in \cite{gell-mann}, which is
expected to play an
important role for our coupled model of vacuum fluctuations as well.

\subsection*{Acknowledgement}

I would like to dedicate this paper to the memory of
Prof.
Friedrich Schl\"{o}gl (1917-2011), who always emphasized the
important role of information in physics.
%who always encouraged me to
%express things in a simple rather than complicated way.

\end{document}